 \documentclass[aip,pop,reprint]{revtex4-1}
 \usepackage{dcolumn}
 \usepackage{bm}
 \usepackage{amsmath}
 
\newcommand{\SP}{\textrm{SP}}
\newcommand{\arctanh}{\textrm{arctanh}}
\newcommand{\D}{\textrm{D}}

\newcommand{\vc}{\mathbf}
\usepackage{graphicx}
\usepackage{color}

\begin{document}

% \preprint{UW-CPTC 09-5}

%\title{Fast reconnection due to plasmoid instability at the Hall scale}
%\title{Plasmoid instability of thin current sheets in Hall-magnetohydrodynamics with an arbitrary guide field}
\title{Reduced magnetohydrodynamic theory of oblique plasmoid instabilities}


\author{S.\ D.\ Baalrud}
%\email[]{sdbaalrud@wisc.edu}
\author{A.\ Bhattacharjee}
\author{Y.-M.\ Huang}

%\thanks{}
\affiliation{Center for Integrated Computation and Analysis of Reconnection and Turbulence, University of New Hampshire, Durham, New Hampshire 03824, USA}

\date{\today}

\begin{abstract}

The three-dimensional nature of plasmoid instabilities is studied using the reduced magnetohydrodynamic equations. For a Harris equilibrium with guide field, represented by $\vc{B}_o = B_{po} \tanh (x/\lambda) \hat{y} + B_{zo} \hat{z}$, a spectrum of modes are unstable at multiple resonant surfaces in the current sheet, rather than just the null surface of the polodial field $B_{yo} (x) = B_{po} \tanh (x/\lambda)$, which is the only resonant surface in 2D or in the absence of a guide field. Here $B_{po}$ is the asymptotic value of the equilibrium poloidal field, $B_{zo}$ is the constant equilibrium guide field, and $\lambda$ is the current sheet width.  Plasmoids on each resonant surface have a unique angle of obliquity $\theta \equiv \arctan(k_z/k_y)$. The resonant surface location for angle $\theta$ is $x_s = - \lambda \arctanh (\tan \theta B_{zo}/B_{po})$, and the existence of a resonant surface requires $|\theta| < \arctan (B_{po} / B_{zo})$.  The most unstable angle is oblique, i.e. $\theta \neq 0$ and $x_s \neq 0$, in the constant-$\psi$ regime, but parallel, i.e. $\theta = 0$ and $x_s = 0$, in the nonconstant-$\psi$ regime. For a fixed angle of obliquity, the most unstable wavenumber lies at the intersection of the constant-$\psi$ and nonconstant-$\psi$ regimes. The growth rate of this mode is $\gamma_{\textrm{max}}/\Gamma_o \simeq S_L^{1/4} (1-\mu^4)^{1/2}$, in which $\Gamma_o = V_A/L$, $V_A$ is the Alfv\'{e}n speed, $L$ is the current sheet length, and $S_L$ is the Lundquist number. The number of plasmoids scales as $N \sim S_L^{3/8} (1-\mu^2)^{-1/4} (1 + \mu^2)^{3/4}$.

\end{abstract}


% 52.35.Vd - Magnetic reconnection
% 52.55.Tn - Ideal and resistive MHD modes; kinetic modes
% 94.30.cp - Physics of the magnetosphere: magnetic reconnection
% 96.60.Iv - Solar physics: magnetic reconnection

%\keywords{}

\maketitle


%%%%%%%%%
\section{Introduction\label{sec:intro}}

\begin{figure}
\includegraphics{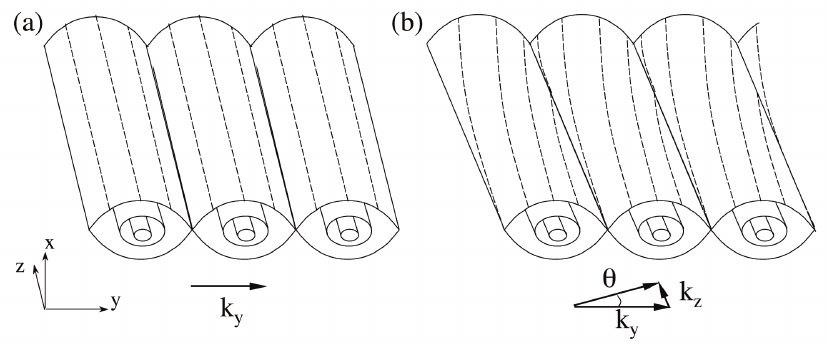}
\caption{Schematic depiction of constant flux surfaces of plasmoids in 3D. Dashed lines designate the guide field direction ($\hat{z}$).  Flux ropes for parallel modes (a) are aligned with the guide field, while for oblique modes (b) they are misaligned by angle $\theta$.} 
\label{fg:fluxt}
\end{figure}

Plasmoid dominated reconnection occurs when a thin current sheet breaks into a chain of secondary islands, or plasmoids, which convect along the reconnection outflow, eventually removing magnetic field from the current sheet.\cite{lour:07,bhat:09,daug:09,samt:09,cass:09,sken:10,ni:10,huan:10,shep:10,uzde:10,huan:11,baal:11,lour:11} Before being ejected, plasmoids may coalesce,\cite{finn:77} and current sheets between plasmoids can excite new generations of plasmoids.\cite{uzde:10,lour:11} This process may be considered turbulent if it can repeat sufficiently many times.\cite{daug:11,lour:09} Since plasmoid dominated reconnection proceeds much faster than the conventional Sweet-Parker\cite{swee:58,park:63} rate, onset of the instability can trigger fast reconnection.  In 3D plasmoids are tube-like in shape, see Fig.~\ref{fg:fluxt}, and are often called flux ropes. There is significant observational evidence for flux ropes in astrophysical reconnection sites including solar flares,\cite{sava:10} and the Earth's magnetopause\cite{russ:78} and magnetotail.\cite{chen:07} Similar tearing instabilities are also important in magnetic confinement fusion experiments.

Magnetohydrodynamic (MHD) theories of tearing instabilities are typically 2D in that they take the guide field direction ($\hat{z}$) to be ignorable.\cite{furt:63,copp:76} For these parallel modes, $k_z=0$, flux ropes are aligned with the guide field as shown in Fig.~\ref{fg:fluxt}a. Oblique tearing modes, with $k_z \neq 0$, are a 3D effect. In tokamak parlance, oblique modes are those with $n \neq 0$, where $n$ is the toroidal mode number. The primary differences between tokamak tearing modes\cite{furt:73} and plasmoid instabilities\cite{lour:07,bhat:09} are the current distributions and boundary conditions. Tearing modes are instabilities of diffuse current distributions, whereas plasmoids are secondary instabilities of thin current sheets. The difference has consequences for how the instabilities scale with resistivity. Boundaries are periodic in both the toroidal and poloidal directions in a tokamak, whereas in astrophysical situations in which current sheets arise boundaries are often open or line-tied. 
%As a consequence, only discrete modes (integer $n$ and $m$) are allowed in a tokamak, whereas a continuum of modes can be unstable for open boundaries. 

It has recently been shown\cite{bhat:09} that the linear plasmoid instability\cite{lour:07} can be related to the conventional tearing mode\cite{furt:63} in a simple way. For a Harris equilibrium\cite{harr:62} without guide field, $\vc{B}_o = B_{po} \tanh (x/\lambda) \hat{y}$, the classical tearing mode growth rate is\cite{copp:76}
\begin{eqnarray}
\gamma \tau_A \sim  \label{eq:coppi}
\left\lbrace \begin{array}{ll}
S^{-3/5} (k\lambda)^{-2/5} (1 - k^2 \lambda^2)^{4/5}, & k\lambda S^{1/4} \gg 1, \\
S^{-1/3} (k\lambda )^{2/3}, & k\lambda S^{1/4} \ll 1 ,
\end{array} \right.
\end{eqnarray}
for the constant-$\psi$ and nonconstant-$\psi$ regimes, respectively. The maximum growth rate occurs at the intersection of the two branches, $k \lambda S^{1/4} \simeq 1$, where $\gamma_{\max} \tau_A \sim S^{-1/2}$. Here, $S=\tau_R/\tau_A$ is the Lundquist number based on the current sheet width, $\tau_R = 4\pi \lambda^2/(\eta c^2)$ is the resistive diffusion time, $\tau_A = \lambda /V_A=\lambda \sqrt{4\pi \rho}/B_{po}$ is the Alfv\'{e}n time, and $k$ is the wavenumber. The recent insight connecting the plasmoid growth rate\cite{lour:07} with Eq.~(\ref{eq:coppi}) was to account for the Lundquist number scaling of the current sheet width.\cite{bhat:09} MHD current sheets obey the Sweet-Parker width $\lambda = \delta_{\SP} = L S_{L}^{-1/2}$, in which $S_L = 4\pi L V_A/(c^2 \eta) = (L/\lambda) S$ is the Lundquist number based on the current sheet length. For a current sheet, the maximum growth rate from Eq.~(\ref{eq:coppi}) is $\gamma_{\max} \simeq S_L^{1/4} V_A/L$, which scales with $S_L$ to a positive exponent, rather than the negative exponent scaling of the most unstable tearing mode. Because current sheets becomes increasingly singular at high $S_L$, the plasmoid growth rate is large for high $S_L$ plasmas such as the solar corona ($S_L \gtrsim 10^{12}$) and fusion experiments ($S_L \gtrsim 10^6$). Plasmoid growth rates that scale with $S_L$ to a positive exponent have also been calculated in the Hall MHD regime.\cite{baal:11}

The present work is motivated by a recent study by Daughton \textit{et al}.,\cite{daug:11} who showed using linear Vlasov and particle-in-cell simulations of the Harris current sheet\cite{harr:62} that the conventional collisionless kinetic theory\cite{gale:86} breaks down for oblique modes. They suggest that this is a failure of asymptotic boundary layer analysis, which is a challenge to analytic theory. Here, we consider the simpler resistive MHD problem and show that, at least within this framework, boundary layer theory can faithfully describe oblique tearing modes. This regime is important in its own right since many reconnection problems of interest are sufficiently collisional that a resistive MHD model is warranted. The spectrum of oblique tearing modes has important consequences for the generation of turbulence by overlapping flux ropes.\cite{daug:11} It is also important when considering whether plasmoids can fill the volume of a current sheet, which is an important assumption in some particle acceleration theories.\cite{drak:06} We find that plasmoids are volume filling, but the angle of obliquity and growth rate depend on the resonant surface location. A numerical study of oblique tearing using Hall MHD without guide field has been presented by Cao and Kan.\cite{cao:91} Huang and Zweibel\cite{huan:09} studies the reduced MHD problem numerically, with guide field, focusing on line-tied boundary conditions. 

%Although tearing instabilities have been an active area of research since their discovery by Furth, Killeen, and Rosenbluth (FKR) in the early 1960's,\cite{furt:63} only recently has their importance for fast reconnection in high Lundquist number plasmas been realized. A possible explanation for this delay may be demonstrated by considering a Harris equilibrium\cite{harr:62} without guide field, $\vc{B}_o = \bar{B}_{yo} \tanh (x/\lambda) \hat{y}$. 

% Since the growth rate scales with $S$ to a negative exponent, tearing modes were expected to be negligible at high Lundquist number. 

%$S_L$ is the appropriate parameter to use in scaling laws because it is the length of a current sheet, rather than the width, that is determined by geometrical constraints.

%The linear tearing mode growth rate has been considered within this context for resistive,\cite{lour:07} and Hall MHD,\cite{baal:11} models.  These theories solved a 2D problem assuming that there is either no guide field, or that there is no variation in the guide field direction. Here we consider the 3D problem within the framework of reduced MHD.\cite{stra:76} Reduced MHD assumes a strong guide field, and long wavelength perturbations in the guide field direction. 

Plasmoid instabilities arise at resonant surfaces of the ideal MHD equations, defined by $\vc{k} \cdot \vc{B}_o = 0$, near which dissipation (resistivity here) allows for reconnection of magnetic field.\cite{furt:63} The primary difference between the 2D and 3D theories is the location of resonant surfaces. For a Harris sheet with guide field, $\vc{B}_o = B_{po} \tanh (x/\lambda) \hat{y} + B_{zo} \hat{z}$, resonant surfaces are located at $x_s = -\lambda \arctanh ( \tan \theta B_{zo}/B_{po})$, where $\theta = \arctan(k_z/k_y)$ is the angle of obliquity. In the conventional 2D theories, either $B_{zo}=0$ or $k_z=0$, in which case there is a single resonant surface corresponding to the null of the sheared field $x_s =0$. In the 3D problem resonant surfaces can be found at any location across the current sheet. Modes at each surface correspond to different angles of obliquity. The angle for a mode at surface location $x$ is $\theta = \arctan [\tanh (-x/\lambda) B_{po}/B_{zo}]$.  Parallel modes ($\theta = 0$) are found at the null surface of the poloidal field and the magnitude of the angle of the mode increases with distance from the null surface. The existence of a resonant surface requires $|\theta | < \arctan (B_{po}/B_{zo})$. We show that for large $k$ (the constant-$\psi$ regime) the most unstable mode is oblique, satisfying $\theta \simeq \pm (B_{po}/B_{zo}) \sqrt{(1 + k^2 \lambda^2)/3}$, and that parallel modes are a local minimum of the growth rate. For small $k$ (the nonconstant-$\psi$ regime), parallel modes are the most unstable, and the growth rate falls off monotonically with $|\theta|$. 
%For a fixed angle of obliquity, the most unstable wavenumber lies at the intersection of the two branches, and this is a parallel mode.
% Far from the null surface, $|x/\lambda| \gtrsim 1$, the angle asymptotes to $\theta \simeq \arctan(\bar{B}_{yo}/B_{zo})$. 

The rest of this paper is organized as follows. Section~\ref{sec:rmhd} describes the reduced MHD equations that form the basis of this analysis. Section~\ref{sec:bl} provides a boundary layer theory for the tearing mode growth rate, and this is used to derive the dispersion relation for the plasmoid instability. These results are compared with direct numerical solutions of the linearized reduced MHD equations in Sec.~\ref{sec:growth}. Section~\ref{sec:num} discusses numerical solutions of the flux and stream functions, which change with the angle of the mode. Section~\ref{sec:sum} provides a summary of the results. 

%%%%%%%%%
\section{Reduced MHD Equations\label{sec:rmhd}}

The reduced MHD equations are based on tokamak ordering\cite{stra:76}
\begin{align}
\partial_x, \partial_y, B_{zo} & \sim \mathcal{O}(1) , \label{eq:tokamak} \\ \nonumber
\partial_z, \partial_t, \psi_o, \phi_o, V_{zo} & \sim \mathcal{O}(\epsilon) , \\ \nonumber
B_{z1} , V_{z1} , \psi_1, \phi_1 & \sim \mathcal{O}(\epsilon^2)  ,
\end{align}
assuming a constant plasma density, a strong, constant, guide field (in the $\hat{z}$ direction here), and that wavelengths in the guide field direction are much longer than in the perpendicular directions.

Applying these approximations, the MHD equation of motion
\begin{equation}
(\partial_t + \vc{V} \cdot \nabla) \vc{V} = \vc{J} \times \vc{B} - \nabla P, \label{eq:eqm}
\end{equation}
Ohm's law
\begin{equation}
\vc{E} + \vc{V} \times \vc{B} = S^{-1} \vc{J}  , \label{eq:ohm}
\end{equation}
and the relevant Maxwell equations $\nabla \cdot \vc{B} = 0$, $\nabla \times \vc{E} = - \partial_t \vc{B}$, and $\nabla \times \vc{B} = \vc{J}$, lead to the reduced MHD equations:\cite{stra:76}
\begin{equation}
\partial_t \Omega + [\Omega, \phi] = [J_z, \psi] +B_z \partial_z J_z  , \label{eq:romega}
\end{equation}
\begin{equation}
\partial_t \psi = B_z \partial_z \phi + [\phi, \psi] + S^{-1} \nabla_\perp^2 \psi + E_o . \label{eq:rpsi}
\end{equation}
Here, the stream function is defined by $\vc{V} = \nabla_\perp \phi \times \hat{z} + V_z \hat{z}$, the flux function by $\vc{B} = \nabla_\perp \psi \times \hat{z} + B_z \hat{z}$, $\Omega \equiv - \nabla_\perp^2 \phi$ is the vorticity, $J_z = -\nabla_\perp^2 \psi$ is the electric current in the $\hat{z}$ direction, $E_o$ and $V_o$ are constants of integration, $\vc{\nabla}_\perp = \partial_x \hat{x} + \partial_y \hat{y}$ is the perpendicular gradient, and $[f,g] = (\nabla f \times \nabla g) \cdot \hat{z}$ is the Poisson bracket. Spatial scales are normalized to the current sheet width ($\tilde{\vc{x}} = \vc{x}/\lambda$), velocities to the Alfv\'{e}n speed ($\tilde{\vc{V}} =\vc{V}/V_A$), time to the Alfv\'{e}n time $\tilde{t} = V_A t/\lambda$, magnetic field to the magnitude of the asymptotic poloidal magnetic field ($\tilde{\vc{B}} = \vc{B}/B_{po}$), and currents by $\tilde{\vc{J}} = \vc{J}/[c B_{po}/(4\pi \lambda)]$. Tildes have been omitted for notational convenience. 

We linearize Eqs.~(\ref{eq:romega}) and (\ref{eq:rpsi}) according to $f = f_o + \delta f$ in which $\psi_o = \psi_o(x)$, $\phi_o = \phi_o(x,y)$, and $B_{zo}$ is constant. We also assume that flow profiles satisfy, $\nabla_\perp^2 \phi_o = 0$, and that the instability growth rate is much larger than the timescale for equilibrium flows, $\partial_x \phi_o, \partial_y \phi_o \ll \gamma$. Perturbations satisfy $\delta f = f_1 (x) \exp[i(k_y y + k_z z) + \gamma t]$. Applying this procedure, and the tokamak ordering (\ref{eq:tokamak}), the linearized form of Eqs.~(\ref{eq:romega}) and (\ref{eq:rpsi}) are:
\begin{equation}
\gamma( \phi_1^{\prime \prime} - k_y^2 \phi_1) = i F (\psi_1^{\prime \prime} - k_y^2 \psi_1) - i F^{\prime \prime} \psi_1 , \label{eq:lphi}
\end{equation}
\begin{equation}
\gamma \psi_1 = i F \phi_1 + S^{-1} (\psi_1^{\prime \prime} - k_y^2 \psi_1) , \label{eq:lpsi}
\end{equation}
in which $F \equiv \vc{k} \cdot \vc{B}_o$ and primes denote $x$ derivatives. Tokamak ordering implies $k = |k_y| [1 + \mathcal{O}(\epsilon)]$ and $k_z/k_y = \tan(\theta) \simeq \theta + \mathcal{O}(\epsilon^3)$, where $\theta \sim \mathcal{O}(\epsilon)$. In the remainder of this work, we adopt the notation $(k,\theta)$ in place of $(k_y, k_z)$. Thus, $F=\vc{k} \cdot \vc{B}_o = k [B_{oy}(x) + \theta B_{oz}]$, which is an $\mathcal{O}(\epsilon)$ quantity.

%%%%%%%%%
\section{Boundary Layer Analysis\label{sec:bl}}

%%%%%%%%%
\subsection{Outer Region}

In the outer region, we assume $S^{-1} \ll \gamma \ll 1$, in which case Eq.~(\ref{eq:lphi}) reduces to the classical ideal MHD outer region of Furth, Killeen, and Rosenbluth (FKR)\cite{furt:63}
\begin{equation}
\psi_1^{\prime \prime} - (k^2 + F^{\prime \prime}/F) \psi_1 = 0 . \label{eq:ideal}
\end{equation}
Equation~(\ref{eq:ideal}) holds everywhere except a small region about the resonant surface, where $F=0$. We follow the conventional boundary layer analysis, which matches the jump in the first derivative of $\psi_1$, denoted the tearing stability index
\begin{equation}
\Delta^\prime \equiv [\psi_1^\prime (x_s^+) - \psi_1^\prime (x_s^-)]/\psi_1(x_s) , \label{eq:dprime}
\end{equation}
in the inner and outer regions. Here $x_s^\pm = \lim_{\epsilon \rightarrow 0} (x_s \pm \epsilon)$ and $\psi_1$ is continuous at $x_s$.  
%Equation~(\ref{eq:ideal}) is the ideal force balance equation for tokamak ordering, but the general solution can be obtained by replacing $k_y$ with the total wavenumber $k$.

\begin{figure}
\includegraphics{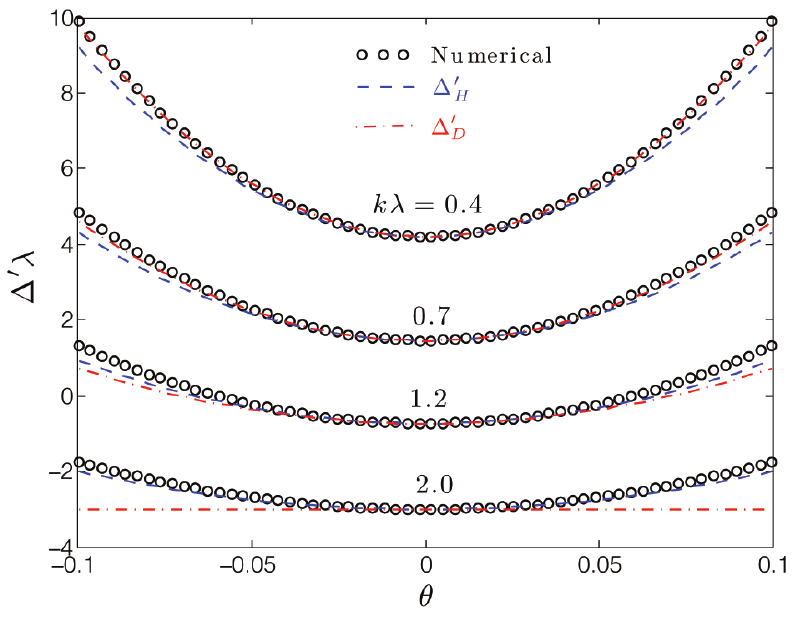}
\caption{(Color online) Angular dependence of the tearing stability index for a Harris equilibrium with $B_{po}/B_{zo} = 0.1$, and four different wavelengths: $k\lambda = 0.4, 0.7, 1.2$ and $2.0$. Black circles show a numerical solution of Eq.~(\ref{eq:ideal}), blue dashed lines the FKR approximation from Eq.~(\ref{eq:harrisdp}), and red dash-dotted lines Daughton's approximation from Eq.~(\ref{eq:daughton}). }
\label{fg:dp}
\end{figure}

\begin{figure}
\includegraphics{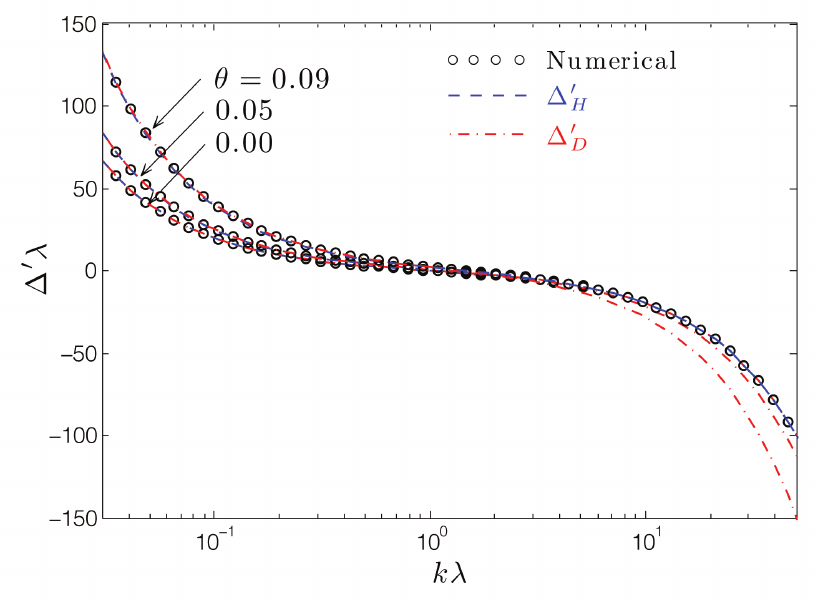}
\caption{(Color online) Wavenumber dependance of the tearing stability index for a Harris equilibrium with $B_{po}/B_{zo} =0.1$, and three values of the angle of obliquity: $\theta = 0.00, 0.05$ and $0.09$ radians. Black circles show a numerical solution of Eq.~(\ref{eq:ideal}), blue dashed lines the FKR approximation from Eq.~(\ref{eq:harrisdp}), and red dash-dotted lines Daughton's approximation from Eq.~(\ref{eq:daughton}). }
\label{fg:dpkscan}
\end{figure}

FKR provides an asymptotic analysis for $\Delta^\prime$ in the large and small $k$ limits. For $k^2 \gg \partial_x^2$, the solution of Eq.~(\ref{eq:ideal}) is $\psi_1 = \psi_1(x_s) \exp(-k |x-x_s|)$, so $\Delta^\prime \rightarrow -2k$ in this limit. For $k^2 \ll \partial_x^2$, FKR show $\Delta^\prime \rightarrow (1/k) [F^\prime(x_s)]^2 (F_{-\infty}^{-2} + F_{\infty}^{-2})$. An approximate solution that captures both the large and small $k$ limits can be obtained by adding the asymptotic solutions
\begin{equation}
\Delta^\prime \simeq (\alpha^2/k) (F_{-\infty}^{-2} + F_{\infty}^{-2}) - 2 k . \label{eq:apdp}
\end{equation}
in which $\alpha \equiv F^\prime (x=x_s) = k B_{oy}^\prime(x=x_s)$.

For the Harris equilibrium with a guide field, $\vc{B}_o = B_{po} \tanh(x) + B_{zo}$, and $x_s = - \lambda \arctanh(\mu)$ where $\mu \equiv k_z B_{zo}/(k_y B_{po}) \simeq \theta B_{zo}/B_{po}$. Thus, $\alpha = k B_{po} (1- \mu^2)$ and Eq.~(\ref{eq:apdp}) yields
\begin{equation}
\Delta_\textrm{H}^\prime \simeq 2 [(1 + \mu^2)/k - k] .  \label{eq:harrisdp}
\end{equation}
Figures~\ref{fg:dp} and \ref{fg:dpkscan} show that Eq.~(\ref{eq:harrisdp}) agrees well with numerical solutions of Eq.~(\ref{eq:ideal}) for the Harris equilibrium. Solutions are shown for $B_{po}/B_{zo} = 0.1$, at fixed wavenumbers varying the angle (Fig.~\ref{fg:dp}), and at fixed angles varying the wavenumber (Fig.~\ref{fg:dpkscan}). Daughton \textit{et al.}\cite{daug:11} have also proposed the solution 
\begin{equation}
\Delta_{\D}^\prime \simeq 2 \biggl( \frac{1}{k} - k \biggr) \biggl[ 1 + \mu^2 \frac{(1-k/2)}{1-k} \biggr]  , \label{eq:daughton}
\end{equation}
for a Harris equilibrium with guide field. Predictions of Eq.~(\ref{eq:daughton}) are also shown in Figs.~\ref{fg:dp} and \ref{fg:dpkscan}.
% Note that the existence of a resonant surface requires $|\theta| \lesssim \bar{B}_{yo}/B_{zo}$. 

For small $k$, Eqs.~(\ref{eq:harrisdp}) and (\ref{eq:daughton}) both asymptote to $\Delta^\prime \rightarrow 2 (1+\mu^2)/k$. However, for large $k$, $\Delta_{\textrm{H}}^\prime \rightarrow -2k$, while $\Delta_{\D}^\prime \rightarrow -k(2+\mu^2)$.  The $\mu$ dependence of the large $k$ limit of Eq.~(\ref{eq:daughton}) is incorrect, as the asymptotic solution of Eq.~(\ref{eq:ideal}) and Fig.~\ref{fg:dpkscan} show. However, Eq.~(\ref{eq:daughton}) provides an excellent approximation for small $k$. This is typically the most interesting case since tearing instability requires $\Delta^\prime > 0$ and $\Delta^\prime$ becomes negative for large $k$. The simple expression (\ref{eq:harrisdp}) provides an adequate approximation for all $k$ and $\theta$, capturing both asymptotic limits. Both results are exact for normal modes ($\theta = 0$).\cite{furt:63} Figure~\ref{fg:dp} shows that an interesting situation can arise for $k \sim 1$, where oblique modes are unstable, $\Delta^\prime (\theta \neq 0) > 0$, but normal modes are stable, $\Delta^\prime (\theta \simeq 0) < 0$. This feature is discussed in detail in Sec.~\ref{sec:growth}. Figures \ref{fg:dp} and \ref{fg:dpkscan} include only $|\theta| < 0.1$, since there is no resonant surface for $|\theta| \geq B_{po}/B_{zo} = 0.1$.

%%%%%%%%
\subsection{Inner Region}

In the inner region, $x - x_s \equiv \xi \ll 1$, we assume $\partial_x^2 = \partial_\xi^2 \gg k_y^2$ and expand $F$ to linear order about the resonant surface: $F \simeq F^\prime (x_s) (x-x_s) \equiv \alpha \xi$. Here, Eqs.~(\ref{eq:lphi}) and (\ref{eq:lpsi}) reduce to
\begin{equation}
\gamma (i \phi_1)^{\prime \prime} = -\alpha \xi (\psi_1 )^{\prime \prime} \label{eq:phi1}
\end{equation}
and
\begin{equation}
\gamma \psi_1 - \alpha \xi (i \phi_1) = S^{-1} \psi_1^{\prime \prime} .  \label{eq:psi1}
\end{equation}
Equations~(\ref{eq:phi1}) and (\ref{eq:psi1}) are the same equations used by Coppi \textit{et al.},\cite{copp:76} to calculate $\Delta^\prime$ in the inner layer. We provide an alternative derivation using a Fourier transform method similar to that developed by Bondeson \textit{et al}.\cite{bond:84} and Porcelli \textit{et al}.\cite{porc:87,fitz:04} 

The fourth-order system of equations (\ref{eq:phi1}) and (\ref{eq:psi1}) has solutions with $\psi_1$ constant, linear, and a solution where $\psi_1$ is an even function of $\xi$ [$\psi_1 (-\xi) = \psi_1 (\xi)$]. We are interested in the last of these. For large $\xi$, Eqs.~(\ref{eq:phi1}) and (\ref{eq:psi1}) reduce to 
\begin{equation}
(\gamma/\alpha)^2 (\psi_1/\xi)^{\prime \prime} + \xi \psi_1^{\prime \prime} = 0
\end{equation}
which has the solution $\psi_1 = a_1 \xi + a_2 \xi \arctan(\alpha \xi /\gamma)$. For large $\xi$, the asymptotic limit of this is
\begin{equation}
\psi_1 \rightarrow A |\xi | + B \biggl( 1 - \frac{\gamma^2}{3 \alpha^2} \frac{1}{\xi^{2}} \biggr) + \mathcal{O}(\xi^{-4}) \label{eq:psi_int}
\end{equation} 
where $A$ and $B$ are constants. Equations (\ref{eq:dprime}) and (\ref{eq:psi_int}) imply 
\begin{equation}
\Delta^\prime = 2A/B . \label{eq:dpab}
\end{equation}

The coefficients $A$ and $B$ can be calculated by matching Eq.~(\ref{eq:psi_int}) with an exact solution of Eqs.~(\ref{eq:phi1}) and (\ref{eq:psi1}). It is convenient to do this matching in a Fourier-transformed space: $\hat{f} (p) = \int_{-\infty}^\infty d\xi\, \exp(-ip\xi) f(\xi)$. Applying this, Eq.~(\ref{eq:psi_int}) can be written
\begin{equation}
\hat{\psi}_1 \rightarrow -\frac{2A}{p^2} + B \biggl( 2\pi \delta(p) + \frac{\pi \gamma^2}{3 \alpha^2} |p| \biggr) + \mathcal{O}(p^3) . \label{eq:psip1}
\end{equation}
The fourth order system of equations (\ref{eq:phi1}) and (\ref{eq:psi1}) can be written as a second order equation for $\hat{\psi}_1$
\begin{equation}
\frac{d}{d\bar{p}} \biggl[ \frac{1}{\bar{p}^2} \frac{d}{d \bar{p}} (\bar{p}^2 \hat{\psi}_1) \biggr] = \Lambda (\Lambda + \bar{p}^2) \hat{\psi}_1  , \label{eq:psip}
\end{equation}
in which $\Lambda \equiv \gamma S^{1/3} \alpha^{-2/3}$, and $\bar{p} \equiv (S \alpha)^{-1/3} p$. The solution of Eq.~(\ref{eq:psip}) is
\begin{equation}
\hat{\psi}_1 = \frac{C_1}{\bar{p}^{3/2}} M_{\nu, 3/4} (\sqrt{\Lambda} \bar{p}^2) + \frac{C_2}{\bar{p}^{3/2}} W_{\nu, 3/4} (\sqrt{\Lambda} \bar{p}^2) , \label{eq:psiwhit}
\end{equation}
in which $M$ and $W$ are the Whittaker-M and W functions, $\nu \equiv - \Lambda^{3/2}/4$, and $C_1$ and $C_2$ are constants. 

The first term of Eq.~(\ref{eq:psiwhit}) diverges for large $\bar{p}$, so we must take $C_1 =0$. For $\bar{p} \ll 1$, 
\begin{equation}
W_{\nu,3/4}/\bar{p}^{3/2} = a (\bar{p}^{-2} + \Lambda^{7/4}/2) + b |\bar{p}| + \mathcal{O}(\bar{p}^2)
\end{equation}
in which $a \equiv \sqrt{\pi} /\lbrace 2 \Lambda^{1/8} \Gamma [(\Lambda^{3/2} + 5)/4] \rbrace$ and $b = 4\sqrt{\pi} \Lambda^{5/8} / \lbrace 3 \Gamma[(\Lambda^{3/2} -1)/4] \rbrace$. Matching the coefficients of Eqs.~(\ref{eq:psip1}) and (\ref{eq:psiwhit}) gives $A= - C_2 (S\alpha)^{2/3} a/2$, and $B = 3C_2 (S\alpha)^{1/3} b /(\pi \Lambda^2)$. With these, Eq.~(\ref{eq:dpab}) provides the dispersion relation
\begin{equation}
\Delta^\prime = - \frac{\pi}{8} (S \alpha)^{1/3} \Lambda^{5/4} \frac{\Gamma [ (\Lambda^{3/2} - 1)/4]}{\Gamma [(\Lambda^{3/2} + 5)/4]}   . \label{eq:coppidp}
\end{equation}
Equation~(\ref{eq:coppidp}) was first obtained by Coppi \textrm{et al}.\cite{copp:76}

Analytically tractable solutions for the growth rate can be obtained from Eq.~(\ref{eq:coppidp}) in the limits $\Lambda \ll 1$ (the constant-$\psi$ regime) 
\begin{equation}
\gamma = \biggl[ \frac{\Gamma(1/4)}{2\pi \Gamma (3/4)} \biggr]^{4/5} S^{-3/5} \alpha^{2/5} \Delta^{\prime 4/5} \label{eq:gcpsi}
\end{equation}
and $\Lambda \rightarrow 1^-$ (the nonconstant-$\psi$ regime) 
\begin{equation}
\gamma = \alpha^{2/3} S^{-1/3} - \frac{2 \sqrt{\pi} \alpha}{3 \Delta^\prime}  . \label{eq:ncpsi}
\end{equation}
The second term in Eq.~(\ref{eq:ncpsi}) causes stabilization at large $k$, but is typically negligible for the most unstable mode.

%%%%%%%%
\section{Linear Growth Rate\label{sec:growth}}
 
The dispersion relation for the linear tearing mode growth rate is obtained by equating Eqs.~(\ref{eq:harrisdp}) and (\ref{eq:coppidp}). These equations can also be used to derive the plasmoid dispersion relation for a Sweet-Parker current sheet, as was done for parallel modes in Ref.~\onlinecite{bhat:09} using the method summarized in Sec.~\ref{sec:intro}.  In terms of unnormalized units, the tearing mode growth rate in the constant-$\psi$ and nonconstant-$\psi$ regimes from Eqs.~(\ref{eq:gcpsi}) and (\ref{eq:ncpsi}) are
\begin{eqnarray}
\gamma \tau_A \sim  \label{eq:growth}
\left\lbrace \begin{array}{ll}
S^{-3/5} (k\lambda)^{-2/5} (1-\mu^2)^{2/5} (1 + \mu^2 - k^2 \lambda^2)^{4/5} \\ 
S^{-1/3} (k\lambda )^{2/3} (1-\mu^2)^{2/3} .
\end{array} \right.
\end{eqnarray}
For a fixed angle of obliquity, the most unstable wavenumber occurs at the intersection of the two regimes, which is $k\lambda \simeq S^{-1/4} (1-\mu^2)^{-1/4} (1 + \mu^2)^{3/4}$. Constant-$\psi$ corresponds to $k\lambda$ larger than this value, and nonconstant-$\psi$ to $k\lambda$ smaller than this value. The leading constant coefficients in Eq.~(\ref{eq:growth}) are $[\Gamma(1/4) /(\pi \Gamma (3/4))]^{4/5} \simeq 0.95$ for the constant-$\psi$ regime, and unity for the nonconstant-$\psi$ regime. Equation~(\ref{eq:growth}) reduces to the classical tearing mode dispersion relation of Eq.~(\ref{eq:coppi}) for normal modes $(\mu = 0)$. The maximum growth rate is $\gamma_{\textrm{max}} \tau_A \simeq S^{-1/2} (1-\mu^4)^{1/2}$. 

\begin{figure}
\includegraphics{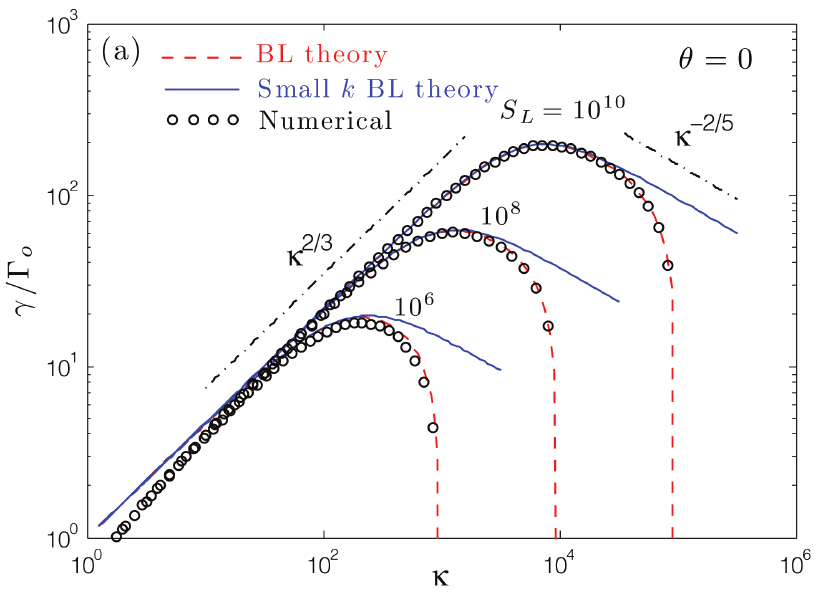}
\includegraphics{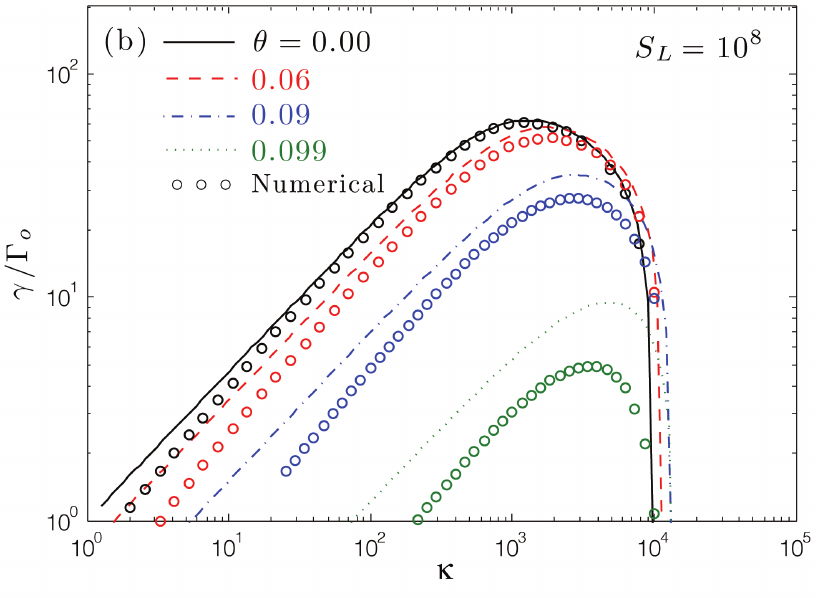}
\caption{(Color online) Wavenumber dependence of the plasmoid growth rate for a Harris equilibrium and $B_{po}/B_{zo} = 0.1$. (a) Normal modes ($\theta = 0$) for three values of the Lundquist number $S_L = 10^6$, $10^8$, and $10^{10}$. Circles show the growth rate from a direct numerical solution of Eqs.~(\ref{eq:lphi}) and (\ref{eq:lpsi}), red dashed lines from the boundary layer theory of Eqs.~(\ref{eq:harrisdp}) and (\ref{eq:coppidp}), and the blue solid line from boundary layer theory using the small $k$ limit of Eq.~(\ref{eq:harrisdp}) [$\Delta^\prime \lambda \simeq 2/(k\lambda)]$ and Eq.~(\ref{eq:coppidp}). (b) Oblique modes with angles $\theta = 0.00, 0.06, 0.09$ and $0.099$ at fixed $S_L = 10^8$. Lines represent the boundary layer theory and circles the numerical solutions.}
\label{fg:mu0}
\end{figure}
%The constant-$\psi$ ($\kappa^{-2/5}$) and nonconstant-$\psi$ ($\kappa^{2/3}$) scalings are also shown.

\begin{figure}
\includegraphics{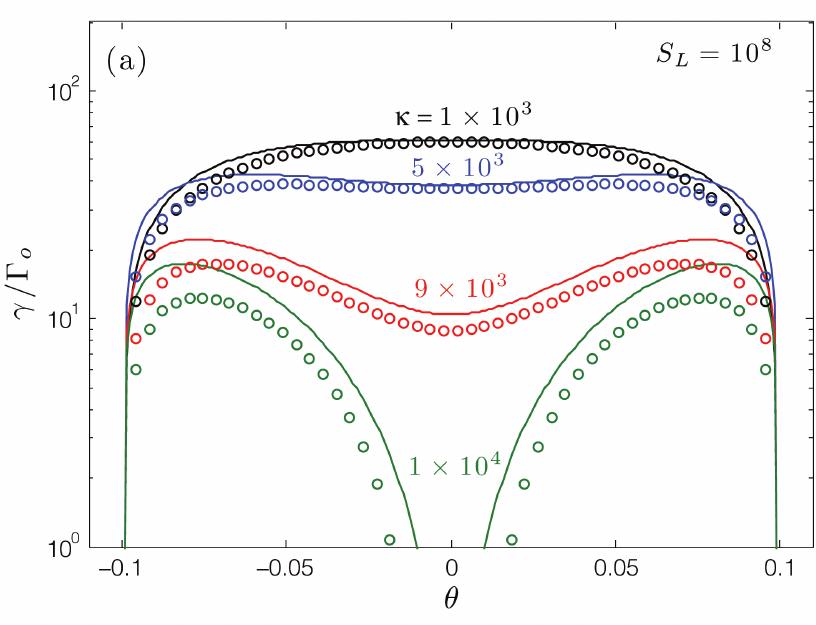}
\includegraphics{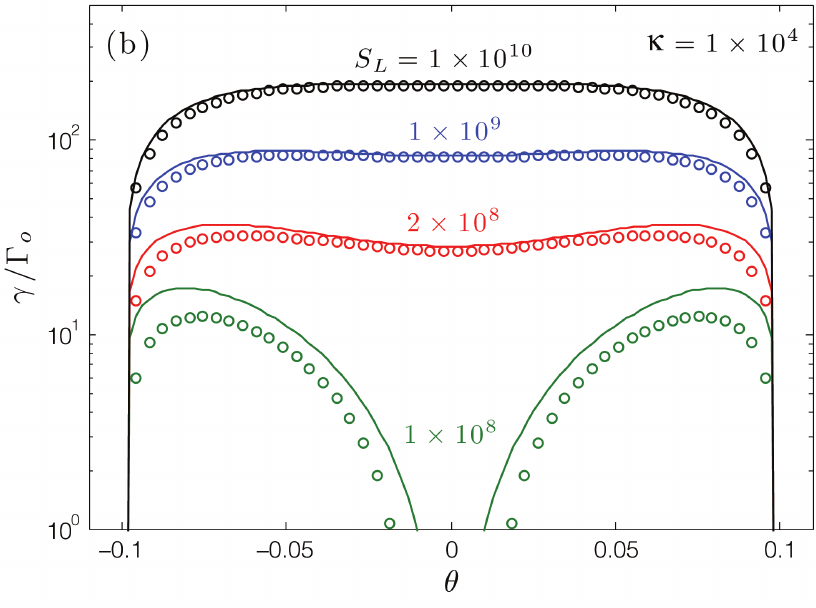}
\caption{(Color online) Angular dependence of the plasmoid growth rate for a Harris equilibrium and $B_{po}/B_{zo} = 0.1$. Lines represent solutions from the boundary layer theory of Eqs.~(\ref{eq:harrisdp}) and (\ref{eq:coppidp}), and circles from direct numerical solutions of Eqs.~(\ref{eq:lphi}) and (\ref{eq:lpsi}). (a) For fixed Lundquist number $S_L = 10^8$, and four values of the wavenumber: $\kappa = 1 \times 10^3, 5 \times 10^3, 9 \times 10^3$, and $1 \times 10^4$.  (b) For fixed wavenumber $\kappa = 1 \times 10^4$ and four values of the Lundquist number $S_L = 1\times 10^8, 2\times 10^8, 1\times 10^9$ and $1 \times 10^{10}$.}
\label{fg:theta_s}
\end{figure}

\begin{figure}
\includegraphics{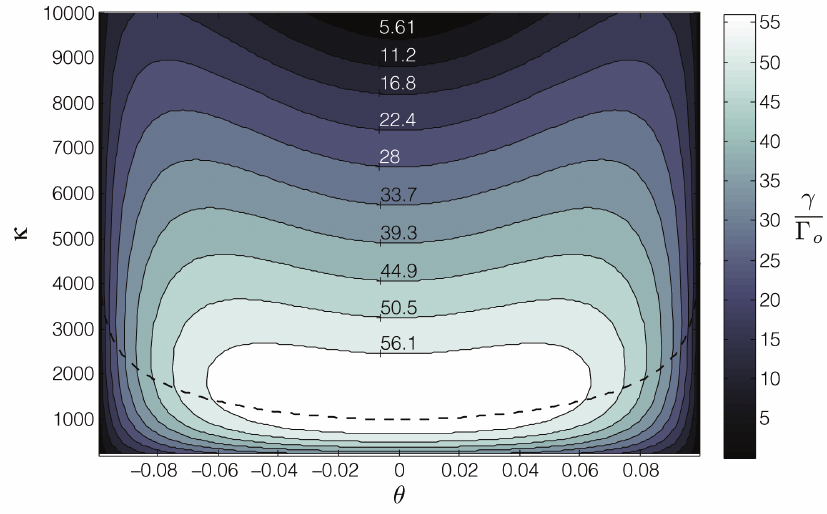}
\caption{(Color online) Contours of the plasmoid growth rate calculated from Eqs.~(\ref{eq:harrisdp}) and (\ref{eq:coppidp}) as a function of wavenumber and angle of obliquity. Here, $S_L = 10^8$ and $B_{po}/B_{zo} = 0.1$. The dashed line shows the estimated boundary between the constant-$\psi$ and nonconstant-$\psi$ regimes from Eq.~(\ref{eq:kappamax}). }
\label{fg:contour}
\end{figure}

Oblique modes ($\mu \neq 0$) are the most unstable in the constant-$\psi$ regime, where the maximum growth rate has angle $\tan \theta = \pm (B_{po}/B_{zo}) \sqrt{(1+k^2 \lambda^2)/3}$. Here, parallel modes are a local minimum in the growth rate. This behavior should be contrasted with that in the nonconstant-$\psi$ regime, where parallel modes ($\mu = 0$) are the most unstable, and the growth rate falls off monotonically for oblique angles. In both regimes, $\mu < 1$ is required for instability, otherwise there is no resonant surface. 

The plasmoid growth rate can be calculated from the tearing mode dispersion relation by taking the equilibrium to be a Sweet-Parker current sheet, which has width $\lambda = \delta_{\SP} = L S_L^{-1/2}$. With this, $S = (\lambda/L) S_L = S_L^{1/2}$, and $\tau_A = 1/(S_L^{1/2} \Gamma_o)$, where $\Gamma_o = V_A/L$. The plasmoid growth rate in the constant-$\psi$ and nonconstant-$\psi$ regimes can then be written
\begin{eqnarray}
\gamma/\Gamma_o \sim  \label{eq:growthsl}
\left\lbrace \begin{array}{ll}
S_L^{2/5} \kappa^{-2/5} (1-\mu^2)^{2/5} (1 + \mu^2 - \kappa^2/S_L)^{4/5} \\ 
\kappa^{2/3} (1-\mu^2)^{2/3}, 
\end{array} \right. 
\end{eqnarray}
in which $\kappa \equiv kL$. 

The most unstable angle in the constant-$\psi$ regime is $\theta \simeq \pm (B_{po}/B_{zo}) \sqrt{(1 + \kappa^2/S_L)/3}$ in these variables. The two regimes meet at 
\begin{equation}
\kappa_{\textrm{max}} \simeq S_L^{3/8} (1-\mu^2)^{-1/4} (1 + \mu^2)^{3/4} \label{eq:kappamax}
\end{equation}
where the maximum growth rate is
\begin{equation}
\gamma_{\textrm{max}}/\Gamma_o \simeq S_L^{1/4} (1-\mu^4)^{1/2}  . \label{eq:gammamax}
\end{equation}
Equation~(\ref{eq:kappamax}) provides an estimate for the number of plasmoids expected to initially arise in an unstable current sheet: $N \simeq \kappa_{\textrm{max}}/(2\pi)$. Parallel modes generate the fewest number of plasmoids, and the plasmoid number increases monotonically with $\theta$. Equations~(\ref{eq:kappamax}) and (\ref{eq:gammamax}) reduce to the results of Loureiro \textit{et al}.\cite{lour:07} for parallel modes. Equation~(\ref{eq:growthsl}) also provides the instability criterion 
\begin{equation}
\kappa < S_L^{1/2} (1+\mu^2)^{1/2} ,  \label{eq:kappacrit}
\end{equation}
for the wavenumber. 

Figures~\ref{fg:mu0} and \ref{fg:theta_s} show the plasmoid growth rate calculated from the boundary layer theory of Eqs.~(\ref{eq:harrisdp}) and (\ref{eq:coppidp}), as well as from a direct numerical solution of the linear reduced MHD equations, (\ref{eq:lphi}) and (\ref{eq:lpsi}). All figures use $B_{po}/B_{zo} = 0.1$. Figure~\ref{fg:mu0}a shows excellent agreement between the theory and numerical results for parallel modes at three values of $S_L$. The $\kappa^{2/3}$ scaling of the nonconstant-$\psi$ regime, and $\kappa^{-2/5}$ scaling of the small $\kappa$ limit ($\kappa^2/S_L \ll 1$) of the constant-$\psi$ regime are also confirmed. The solid lines in Fig.~\ref{fg:mu0}a are obtained using the small $k\lambda$ limit of Eq.~(\ref{eq:harrisdp}) [$\Delta^\prime \lambda = 2/(k\lambda)$], which extends the $\kappa^{-2/5}$ scaling beyond its region of validity. This is the limit assumed in Ref.~\onlinecite{lour:07}, and is shown for comparison. Although the growth rate falls off rapidly for $\kappa > \kappa_{\max}$, the $\kappa^{-2/5}$ scaling holds near the intersection with the nonconstant-$\psi$ regime. The maximum growth rate obtained from the intersection of the nonconstant-$\psi$ regime and the small $\kappa$ limit of the constant-$\psi$ regime provides a good approximation of the full analytic, and numerical, results. 
% However, the domain of $\kappa$ for which the constant-$\psi$ approximation holds is very narrow. 

Figure~\ref{fg:mu0}b shows results for fixed $S_L = 10^8$, and different values of the angle of obliquity. Here, too, the growth rate calculated with boundary layer theory compares well with the numerical results. The agreement becomes less favorable for angles very close to the cutoff angle $|\theta_{\textrm{max}}| \simeq B_{po}/B_{zo} = 0.1$. Figure~\ref{fg:mu0}b shows that parallel modes are the most unstable in the nonconstant-$\psi$ regime. The constant-$\psi$ regime is found at $\kappa \simeq 10^4$, where oblique modes have larger growth rates than parallel modes. The maximum wavenumber for instability from Eq.~(\ref{eq:kappacrit}) shows excellent agreement for all Lundquist numbers and angles shown in Fig~\ref{fg:mu0}.

Figure~\ref{fg:theta_s} again shows close agreement between boundary layer theory and numerically calculated growth rates. Figure~\ref{fg:theta_s}a shows the angular dependence of the growth rate for fixed Lundquist number, $S_L=10^8$, and four values of the wavenumber, whereas Fig.~\ref{fg:theta_s}b fixes the wavenumber, and varies the Lundquist number. In both cases, a transition between the two regimes of the instability are evident. At small $\kappa$, or large $S_L$, parallel modes are most unstable. Here, the growth rate decreases monotonically with $|\theta|$ until the stability threshold at $|\theta| \simeq B_{po}/B_{zo} = 0.1$ is reached, which is indicative of the nonconstant-$\psi$ regime. Modes at larger $\kappa$, or smaller $S_L$, are most unstable at oblique angles. Here, the most unstable angle agrees with the prediction of the constant-$\psi$ regime: $\theta \simeq \pm (B_{po}/B_{zo})\sqrt{(1+\kappa^2/S_L)/3}$. 

A contour plot of the growth rate is shown in Fig.~\ref{fg:contour} for a range of wavenumbers near the peak growth rate and the entire domain of unstable angles. Here the Lundquist number is fixed at $S_L = 10^8$ and the growth rate was calculated from the boundary layer theory using Eqs.~(\ref{eq:harrisdp}) and (\ref{eq:coppidp}). Again, the angular dependence of modes in the constant-$\psi$ regime is evident. Here, a linear scale has been used for $\kappa$, which allows better resolution of the constant-$\psi$ regime. The dashed line shows the estimated $\kappa$ at the maximum growth rate from Eq.~(\ref{eq:kappamax}), which corresponds to the boundary between the constant-$\psi$ (above the dashed line) and nonconstant-$\psi$ (below the dashed line) regimes. 

Oblique modes in the constant-$\psi$ regime of Eq.~(\ref{eq:growth}) are analogous to $n \geq 1$ tearing modes in a tokamak. In fact, all tearing modes in tokamaks are oblique since the presence of $n=0$ modes would require an infinite safety factor (unless $m=0$). Furth, Rutherford, and Selberg considered $n=1$, constant-$\psi$ tearing modes in a periodic cylinder.\cite{furt:73} In cylindrical geometry, the linearization becomes $f_1 (x) \exp [i (k_y y +k_z z)] \rightarrow f_1(r) \exp[i (k z + m \vartheta)]$, where $\vartheta$ is the poloidal direction and $k = n/R$ is quantized according to the tokamak major radius $R$. In this case, the angle of obliquity is $\tan \theta = k_z/k_y \rightarrow r_s n/(Rm)$, where $r_s$ is the minor radial location of the resonant surface. In terms of the safety factor, $q = |rB_z/(RB_\vartheta)|$, $q(r_s) = m/n$, and
\begin{equation}
\tan \theta = \frac{r_s/R}{q(r_s)} .
\end{equation} 
The angles of obliquity are small for large aspect ratio tokamaks, and increase with the ratio $n/m$ simply due to geometry; flux tubes with higher $n$ numbers, or lower $m$ numbers, must travel farther poloidally in a toroidal transit.  The existence of a resonant surface, and the resulting growth rate, both depend on the current profile, which is significantly different in tokamaks than the Harris equilibrium we have assumed above. The effect of periodic boundary conditions on the above analysis would simply require quantization of $k_y$ and $k_z$ according to the length of the current sheet $L$ and the domain size in the guide field direction. In this case, $\kappa \simeq 2\pi m$ and Fig.~\ref{fg:mu0} shows that there are typically thousands of islands in a chain for the current sheet lengths of interest. 

%%%%%%%%
\section{Numerical Eigenmode Solutions\label{sec:num}}

\begin{figure}
\includegraphics{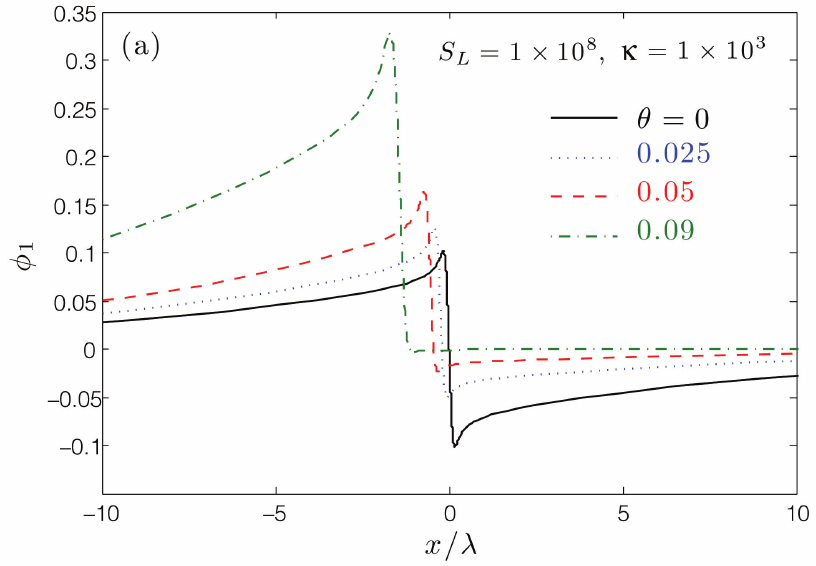}
\includegraphics{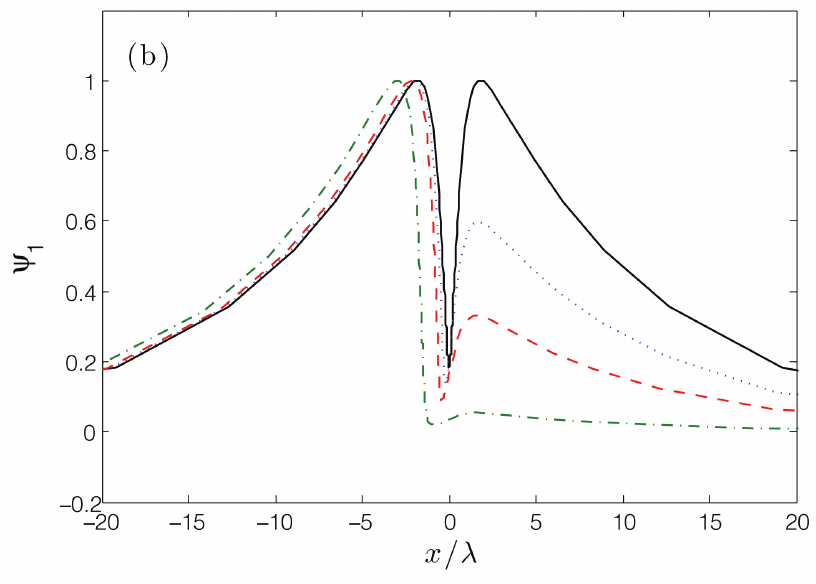}
\caption{(Color online) Perturbed stream function $\phi_1$ (a) and flux function $\psi_1$ (b) for $S_L = 1 \times 10^8$, $\kappa = 1 \times 10^3$, and four values of the angle of obliquity: $\theta = 0, 0.025, 0.05$, and $0.09$ radians. These eigenfunctions were obtained from a direct numerical solution of Eqs.~(\ref{eq:lphi}) and (\ref{eq:lpsi}). }
\label{fg:e_fun}
\end{figure}

\begin{figure*}
\includegraphics{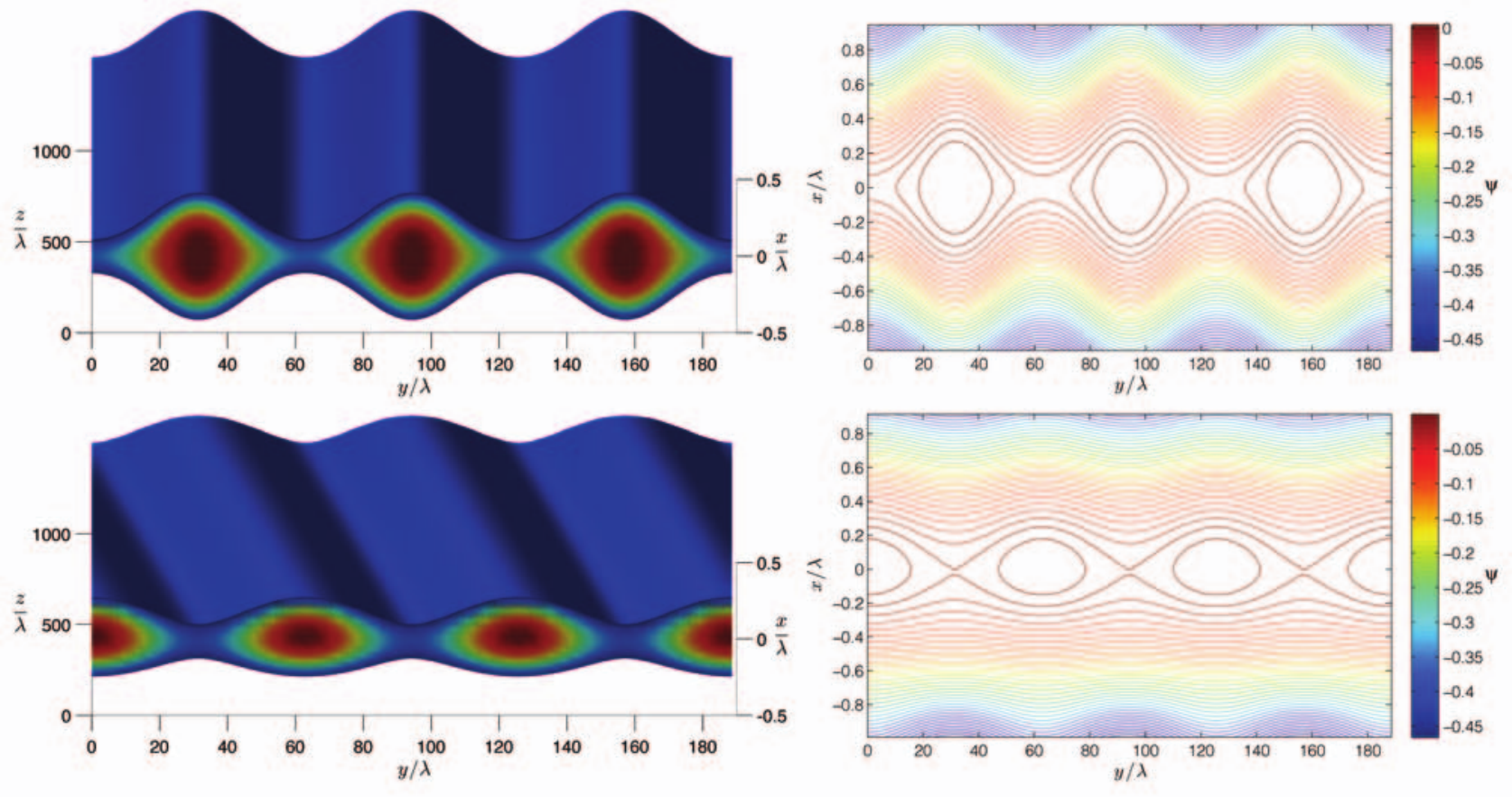}
\caption{ (Color online)  Three-dimensional flux ropes obtained from isosurfaces of the flux function, along with corresponding magnetic islands obtained from 2D cuts in the $x-y$ plane. The $\hat{x}$ direction is vertical and $\hat{z}$ direction mostly into the page in the figure. Here, $\kappa = 10^3$, $S_L = 10^8$, and $B_{po}/B_{zo} = 0.1$. The top row is a parallel mode, where $\theta = 0$, and the bottom row is an oblique mode, where $\theta = 0.06$ radians. The colorbar corresponds to colors in the 2D plots, and represents values of constant flux normalized by $B_{po} \lambda$.}
\label{fg:iso}
\end{figure*}

Figure~\ref{fg:e_fun} shows numerical solutions of the perturbed flux and stream functions for fixed $S_L = 1\times 10^8$, $\kappa = 1 \times 10^3$, and four angles of obliquity: $\theta = 0, 0.025, 0.05$, and $0.09$. As the angle of obliquity increases, two significant changes occur: the eigenfunction center shifts, and symmetries of the eigenfunctions are lost. The shift in eigenfunction center corresponds to the shift of the resonant surface: $x_s/\lambda = - \arctanh(\mu) \simeq - \arctanh(\theta B_{zo}/B_{po})$. For the plotted angles, $\theta = 0.00, 0.025, 0.05$, and $0.09$, the predicted resonant surface locations are $x_s/\lambda = 0, -0.025, -0.055$, and $-1.47$, respectively. The gradient of the perturbed stream function in Fig.~\ref{fg:e_fun}a decreases for $x>x_s$, but grows for $x<x_s$, as the angle of obliquity increases. Likewise, the gradient of the perturbed flux function in Fig.~\ref{fg:e_fun}b decreases for $x>x_s$, but remains nearly constant on the $x<x_s$ side. 
%Qualitatively, this suggests that the tearing stability index, Eq.~(\ref{eq:dprime}), should increase with increasing angle of obliquity. This idea is consistent with the outer region solutions shown in Figs.~\ref{fg:dp} and \ref{fg:dpkscan} that were used in the boundary layer analysis. 

Figure~\ref{fg:iso} shows constant flux surfaces, which generate flux ropes in 3D, for a parallel mode ($\theta = 0$ in the top row) and an oblique mode ($\theta = 0.06$ radians in the bottom row). Here, the parameters $\kappa = 10^3$ and $S_L = 10^8$ have been chosen. Also shown in the right column are 2D cuts of the same data, at $z=0$, showing magnetic islands. The total flux function is used, which is the sum of the equilibrium component, and the perturbed component multiplied by a constant amplitude: $\psi = \psi_o + a \psi_1$. For the values in Fig.~\ref{fg:iso}, $\psi$ is normalized to $B_{po} \lambda$ and $a$ is chosen to be $0.01$. The equilibrium component, $\psi_o$, is obtained from the definition $\vc{B}_o = \nabla_\perp \psi_o \times \hat{z} + B_{zo}$. This implies $d\psi_o/dx = - B_{po} \tanh (x/\lambda)$ for the Harris sheet, so
\begin{equation}
\psi_o = - B_{po} \lambda \ln[\cosh(x/\lambda)] ,
\end{equation}
is used for $\psi_o$.

The flux ropes shown in Fig.~\ref{fg:iso} correspond with the qualitative expectation from Fig.~\ref{fg:fluxt}. For parallel modes, they are straight and uniform in the $\hat{z}$ direction, which is consistent with taking $z$ to be an ignorable direction in the conventional theory. Here, the 2D cuts of magnetic islands are symmetric in $x$ about the resonant surface $x=x_s=0$. For oblique modes, the flux tubes are shifted by angle $\theta$ from the normal in the $z$ direction. Here, the 2D cuts of magnetic islands do not possess the $x$ symmetry of parallel modes, having a slightly shallower gradient for $x>x_s$ than for $x<x_s$. The resonant surface is also shifted slightly $x_s/\lambda \simeq -0.055$.

%%%%%%%%
\section{Summary\label{sec:sum}}

Oblique plasmoid instabilities were analyzed within the context of the reduced MHD approximation. This required accounting for the 3D effect of wave variations in the guide field direction, which is the ignorable direction in the 2D theory. An important difference between the 2D and 3D theories is the location of the resonant surfaces (where $F = \vc{k} \cdot \vc{B}_o=0$). Considering a Harris equilibrium with guide field, this can take any value in 3D: $x_s = -\lambda \arctanh [k_z B_{zo}/(k_y B_{po})]$. In the 2D case, $k_z = 0$ and the resonant surface is always the null surface of the poloidal equilibrium field. The same is true in the 3D case only if there is no guide field present. The boundary layer analysis changed primarily in the outer region, where $\Delta^\prime$ depends on the angle of obliquity [see Eq.~(\ref{eq:apdp})]. The approximate $\Delta^\prime$ expression we obtained for a Harris equilibrium was shown to compare favorably with a numerical solution of the ideal MHD force balance. In the inner region, the angle of obliquity only entered in locating the resonant surface when evaluating $F^\prime(x_s)$. 

We found that unstable modes are confined to small angles $|\theta | \lesssim \arctan (B_{po}/B_{zo})$. In the constant-$\psi$ regime, the most unstable tearing mode is an oblique mode $|\theta| \simeq (B_{po}/B_{zo}) \sqrt{(1+k^2 \lambda^2)/3} \neq 0$. In the nonconstant-$\psi$ regime, the most unstable tearing mode is a parallel mode ($\theta = 0$). The most unstable wavenumber is located at the intersection of these two regimes. The growth rate for this wavenumber is largest in the parallel direction. By choosing an appropriate wavenumber at fixed Lundquist number (or vice versa) it was shown that a situation can arise in which only oblique modes are unstable. The boundary layer theory was shown to compare well with numerical solutions of the linear reduced MHD equations. 

%These results may have important implications for the nonlinear evolution of plasmoid dominated reconnection in 3D. As flux ropes with different $k_z$ (angles of obliquity) grow, they interact with one another 

\begin{acknowledgments}

The authors thank Dr.\ Will Fox, Dr.\ Robert L.\ Dewar, Dr.\ Bill Daughton, and Dr.\ Carl Sovinec for helpful discussions. This research was supported in part by an appointment to the U.S. Department of Energy Fusion Energy Postdoctoral Research Program administered by the Oak Ridge Institute for Science and Education (S.D.B.), and DOE Grant No.\ DE-FG02-07ER46372, NSF Grant Nos.\ ATM-0802727, ATM-0903915, and AGS-0962698, and NASA Grant Nos. NNX09AJ86G and NNX10AC04G.

\end{acknowledgments}

\bibliography{refs.bib}

\end{document}